\documentclass[twocolumn,prl,superscriptaddress]{revtex4}
\usepackage{graphicx}% Include figure files
\usepackage{dcolumn}% Align table columns on decimal point
\usepackage{amsmath,amsthm,amssymb}

\makeatletter
\newlength{\mySpaceUnder}
\newlength{\mySpaceOver}
\renewcommand\section{\@startsection {section}{1}{\z@}%
                                   {\mySpaceOver}%
                                   {\mySpaceUnder}%
                                   {\normalfont\small\bfseries\centering}}

\renewcommand\subsection{\@startsection {subsection}{2}{\z@}%
                                   {\mySpaceOver}%
                                   {\mySpaceUnder}%
                                   {\normalfont\small\bfseries\centering}}
\makeatother

\begin{document}
%

%\preprint{APS/123-QED}

\title{Nonlinear perturbations of the Kaluza-Klein monopole}

\author{Piotr Bizo\'n}
 \affiliation{M.~Smoluchowski~Institute~of~Physics,~Jagiellonian~University,~Krak\'ow,~Poland}
\affiliation{Max-Planck-Institut f\"ur Gravitationsphysik,
Albert-Einstein-Institut, Golm, Germany}
\author{Tadeusz Chmaj}
\affiliation{H. Niewodniczanski Institute of Nuclear Physics,
Krak\'ow, Poland}
\author{Gary  Gibbons}
 \affiliation{Department of Applied Mathematics and Theoretical Physics, Cambridge University, Cambridge, UK}
\date{\today}
\begin{abstract}
We consider the nonlinear stability of the Kaluza-Klein monopole
viewed as the static solution of the five dimensional vacuum
Einstein equations. Using both numerical and analytical methods we
give evidence that the Kaluza-Klein monopole is asymptotically
stable within the cohomogeneity-two biaxial Bianchi IX ansatz
recently introduced in \cite{bcs}. We also show that for
sufficiently large perturbations the Kaluza-Klein monopole loses
stability and collapses to a Kaluza-Klein black hole. The relevance
of our results for the stability of BPS states in M/String theory is
briefly discussed.
\end{abstract}

\maketitle \addtolength{\textheight}{.05in}
\addtolength{\footskip}{-.05in}
\section{Introduction}
The Taub-NUT  instanton is the complete  Riemannian Ricci-flat
4-manifold with the metric~\cite{h}
\begin{eqnarray}\label{inst}
ds^2&=
\left(1+\dfrac{m}{\rho}\right)d\rho^2+\left(1+\dfrac{m}{\rho}\right)\rho^2
(d\theta^2+\sin^2{\theta} d\phi^2)\nonumber \\&+m^2
\left(1+\dfrac{m}{\rho}\right)^{-1} (d\psi+\cos{\theta} d\phi)^2\,,
\end{eqnarray}
where $\rho \geq 0$, $0\leq \theta \leq \pi, 0\leq \phi \leq 2\pi$,
$0\leq \psi \leq 4\pi$ and $m>0$ is a free parameter which sets the
scale. The Taub-NUT instanton is topologically $R^4$ with the
surfaces of constant $\rho$ being the squashed three-spheres.
Although complete, the metric (\ref{inst}) is not asymptotically
Euclidean because the $S^1$ fibers approach a constant circumference
at infinity (rather than growing like $\rho$). This kind of
asymptotic behavior is called   asymptotically locally flat.

By adding the trivial term $-dt^2$ one obtains from (\ref{inst}) the
regular static solution of the five dimensional vacuum Einstein
equations
\begin{eqnarray}\label{tn}
\!\!\!&ds^2\!=\!-dt^2 \!+ \! \left(1+\dfrac{m}{\rho}\right)d\rho^2\!
+\! \left(1+\dfrac{m}{\rho}\right)\rho^2 (d\theta^2+\sin^2{\theta}
d\phi^2)\nonumber \\&+m^2 \left(1+\dfrac{m}{\rho}\right)^{-1}\!
(d\psi+\cos{\theta} d\phi)^2\,.
\end{eqnarray}
This metric has been used in the past to obtain the four-dimensional
magnetic monopole via a Kaluza-Klein reduction along the coordinate
$\psi$ \cite{gps} and since then it is usually called the
Kaluza-Klein (KK) monopole.
 To the best of our knowledge, the role of the KK monopole in the dynamics of non-asymptotically
  flat initial data, and in particular its nonlinear stability, has not been studied, probably because this problem appeared to require an
analysis of the full
  five dimensional Einstein equations which is beyond currently available
  numerical tools, let alone the analytic ones. This  situation
  has changed recently with the introduction of a new
  cohomogeneity-two symmetry reduction
  of the five dimensional vacuum Einstein equations (referred to below as the BCS ansatz \cite{bcs})
  which provided a framework for investigating the dynamics in a simple $1+1$
  dimensional setting. Since the KK monopole falls within
  the BCS ansatz, the question of its nonlinear
  stability now becomes tractable and is the subject~of~this~paper.

  The KK monopole plays an important role in M/String theory. For
  example,
its metric product with six flat Euclidean dimensions gives a
Ricci-flat eleven dimensional Lorentzian  metric, which when reduced
to ten spacetime dimensions may be interpreted as a D6-brane
solution of Type IIA String theory~\cite{t}. As such, it is an
example of what is called a supersymmetric or BPS solution. BPS
solutions of supergravity theories are widely believed by many
string theorists to be absolutely stable, both classically and
quantum mechanically. However until now, this belief
 has never been tested in a fully
non-linear setting. As we shall see, our results show that the
classical stablity is not absolute since for sufficiently large
perturbations a gravitational collapse to a black hole is possible.
The implications of our results for  BPS states will be discussed
briefly in the conclusions.

\section{Preliminaries}
  The BCS ansatz
  has the form \cite{bcs}
\begin{eqnarray}\label{bcs}
    &ds^2= - A e^{-2\delta} dt^2 + A^{-1} dr^2 + \frac{1}{4} r^2
    e^{2B} (d\theta^2+\sin^2{\theta} d\phi^2)\nonumber \\
     &+ \frac{1}{4} r^2 e^{-4B} (d\psi+\cos{\theta} d\phi)^2\,,
\end{eqnarray}
where $A$, $\delta$, and $B$ are functions of $t$ and $r$. In
contrast to spherical symmetry, this ansatz possesses a dynamical
degree of freedom, the field $B$, which measures the deformation of
sphericity of the angular part of the metric.
 Substituting the metric
(\ref{bcs}) into the vacuum Einstein equations one gets the
following system of equations
\begin{subequations}
\begin{eqnarray}\label{mom}
 \!\! A'\! \!&= &\!\!- \!\!\frac{2 A}{r}\! +\!\frac{2}{3r} \!\left(\!4 e^{-2B}\!\! -
 \!\!
  e^{-8B}\!\right)\!-\!
 2r
 \left( e^{2\delta} A^{-1} {\dot B}^2\!\! + \!\!A {B'}^2\right)\,\,\,\,\\
  \!\!  \dot A \!\!&=&\!\! - 4 r A \dot B B'\,, \\
\!\! \delta' \!\!& =& \!\!- 2 r \left(e^{2\delta} A^{-2}{\dot B}^2 +
    B'^2\right)\,,
\end{eqnarray}
\begin{equation}\label{wave}
\left(e^{\delta} A^{-1} r^3 {\dot B}\right)^{\cdot} -
\left(e^{-\delta} A r^3 B'\right)' + \frac{4}{3} e^{-\delta} r
\left(e^{-2B}-e^{-8B}\right)=0\,,
\end{equation}
\end{subequations}
    where primes and dots denote derivatives with respect to $r$ and
$t$, respectively. The initial value problem for this system was
investigated in \cite{bcs} in the case of asymptotically flat
initial data. Here we consider asymptotically locally flat
solutions.

The KK monopole (\ref{tn}) expressed in terms of the ansatz
(\ref{bcs}) takes the form
\begin{equation}\label{tn_bcs}
B_0=\frac{1}{3}\ln\left(1+\dfrac{\rho}{m}\right),\,
A_0=\frac{(1+\dfrac{4\rho}{3m})^2}{(1+\dfrac{\rho}{m})^{8/3}},\,
e^{2\delta_0}=A_0\,,
\end{equation}
where
\begin{equation}\label{rho}
 r=2 m^{1/2} \rho^{1/2}\left(1+\dfrac{\rho}{m}\right)^{1/6}\,.
 \end{equation}
Note that scaling invariance excludes the existence of nontrivial
solutions which are everywhere regular \emph{and} asymptotically
flat. The KK monopole is the unique nontrivial regular static
solution of the system (4). To see this note that static solutions
which are analytic at the origin form a one-parameter family with
the following asymptotic behavior for $r \rightarrow 0$
\begin{equation}\label{bc}
 B(r) \sim b r^2, \quad A(r) \sim 1-4 b^2 r^4, \quad \delta \sim -2
 b^2 r^4\,,
\end{equation}
where $b$ is the shooting parameter. Since the system (4) is scale
invariant, it follows from the uniqueness of solutions of ordinary
differential equations that, up to scaling, there is only one static
solution which is analytic at the origin. It is easy to verify that
the solution (\ref{tn_bcs}) satisfies the regularity conditions
(\ref{bc}) with $b=1/(12 m^2)$. For convenience, hereafter we choose
units in which $m=~1$.
\section{Linear stability}
 Before discussing numerical results, we first demonstrate
that the KK monopole is linearly stable within our ansatz. To this
end, following the standard procedure we seek solutions of the
system (4) in the form
\begin{eqnarray}\label{pert}
    B(t,r)&=&B_0(\rho)+B_1(t,\rho), \quad
    A(t,r)=A_0(\rho)+A_1(t,\rho), \nonumber \\
    \delta(t,r) &=&\delta_0(\rho)+\delta_1(t,\rho)\,,
\end{eqnarray}
where $\rho$ is a function of $r$ determined implicitly by
(\ref{rho}). Substituting the expansion (\ref{pert}) into the system
(4), linearizing and separating the time dependence
$B_1(t,\rho)=\exp(-i \lambda t) v_{\lambda} (\rho)$, after a
straightforward but tedious calculation we get the eigenvalue
equation for the spectrum of small perturbations around the KK
monopole
\begin{eqnarray}\label{eigen}
\mathcal{A} v_{\lambda} &=&\lambda^2 v_{\lambda}, \\
 \mathcal{A}&=& -\frac{1}{\rho
(1+\rho)}\frac{d}{d\rho}\left(\rho^2\frac{d }{d \rho}\right) +
\frac{18}{\rho (1+\rho) (3+4\rho)^2}\,. \nonumber
\end{eqnarray}
Although this equation cannot be solved analytically in general,
there is an explicit solution for $\lambda=0$
\begin{equation}\label{zeromode}
    v_0 = \frac{\rho}{3+4\rho}.
\end{equation}
This is the zero mode corresponding to the scaling freedom  - it can
be obtained by the action of the scaling generator $r\frac{d}{dr}$
on $B_0(\rho(r))$. The zero mode (\ref{zeromode}) has no zeros which
implies by the standard Sturm-Liouville theory that the operator
$\mathcal{A}$  has no negative eigenvalues. Thus, the KK monopole is
linearly stable within our ansatz. Note that the zero mode is not a
genuine eigenfunction because it is not square integrable. More
precisely, on the Hilbert space $L^2\left([0,\infty), \rho(1+\rho)
d\rho\right)$ the self-adjoint operator $\mathcal{A}$ has the purely
continuous spectrum $\lambda^2\in[0,\infty)$.

Nota bene, the Taub-NUT instanton (\ref{inst}) is known to be
linearly stable against transverse traceless perturbations. This
follows from the work of Hawking and Pope~\cite{hp}, who showed,
using two linearly independent covariantly constant spinors which
the Taub-NUT metric admits, that  the spectrum of the Lichnerowicz
operator, which governs the linearized transverse traceless
perturbations, is the same (apart from zero modes) as for the scalar
Laplacian, and therefore non-negative.
 \section{Numerical results}
  Using a fourth-order accurate finite
 difference code
 we have solved the system
(4) numerically for several families of regular initial data which
represent various perturbations of the KK monopole. The overall
picture does not depend on the specific choice of a family. The
results shown below were produced for initial data of the form
(using the momentum variable $P=e^{\delta} A^{-1} \dot B$)
\begin{equation}\label{idata}
 \!   B(0,r)=B_0(\rho(r)), \,\,\, P(0,r)=p \left(\frac{r}{R}\right)^4
    e^{-\frac{(r-R)^4}{s^4}},
\end{equation}
where the amplitude $p$ was varied and the parameters $R$ and $s$
were kept fixed. Note that although the perturbation  (\ref{idata})
is exponentially localized, the induced perturbation of the function
$A$ has a $1/r^2$ tail as follows from the hamiltonian constraint
(4a).
\subsection{Small  perturbations}
We have found that for small perturbations, that is for small values
of the control parameter $p$, the perturbation is scattered off to
infinity and the solution returns to equilibrium, i.e., it settles
down to the KK monopole with the \emph{same} parameter $m$. This is
shown in Fig.~1. One might wonder why the parameter $m$ does not
change under perturbation. The reason is simple: the perturbations
considered by us have finite energy while a change of $m$ would
require infinite energy. Here by energy we mean the energy measured
with respect to a given KK reference background as described by
Deser and Soldate~\cite{ds}. This energy is determined by the
\emph{next} to leading order term in the asymptotic expansion at
spatial infinity.

The qualitative picture of pointwise convergence to the KK monopole
is shown in Fig.~2. The details of this process, in particular the
role of  quasinormal modes and the rate of decay of a tail, will be
pursued elsewhere.

\begin{figure}[h]
\centering
\includegraphics[height=0.4\textwidth,width=0.48\textwidth]{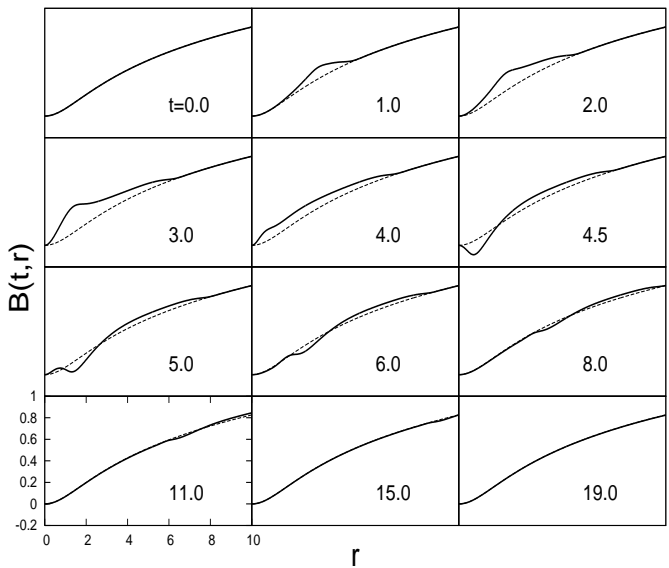}
\caption{\small{Asymptotic stability of the Kaluza-Klein monopole.
For initial data (\ref{idata}) with a small amplitude ($p=0.1, R=3,
s=1$) we plot a series of snapshots of the function $B(t,r)$ (where
$t$ is central proper time). The dashed line shows the unperturbed
KK monopole. During the evolution the excess energy of the
perturbation is clearly seen to be radiated away to infinity and the
solution returns to equilibrium.} }\label{fig1}
\end{figure}
\begin{figure}
\centering
\includegraphics[width=0.48\textwidth]{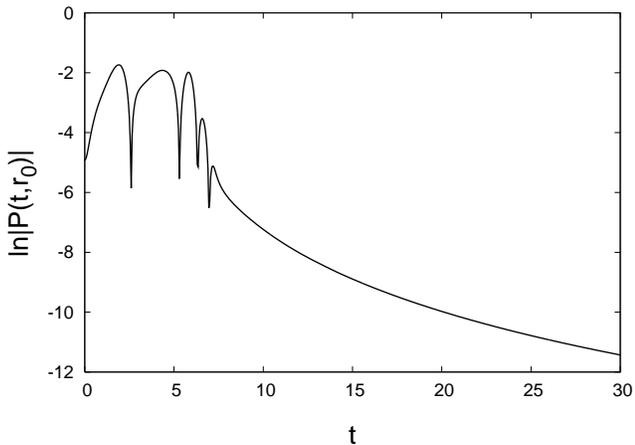}
\caption{\small{The convergence to the Kaluza-Klein monopole. For
the same initial data as in Fig.~1 we plot the time series
$\ln|P(t,r_0)|$ at $r_0=2$. The asymptotic power-law tail is seen at
late times. } }\label{fig2}
\end{figure}

 We have found that a
sufficiently strong kick may destabilize the KK monopole. In order
to understand what is the fate of a strongly perturbed KK monopole
we need to make a digression.
\subsection{Kaluza-Klein black holes}
 The KK monopole (\ref{tn}) is in fact a special case of the
two-parameter family of solutions of the vacuum Einstein equations
in five dimensions \cite{gw}
\begin{eqnarray}\label{tnbh}
& ds^2=-\dfrac{\rho}{\Delta} dt^2+ \dfrac{\Sigma}{\rho}
d\rho^2+\Delta \Sigma (d\theta^2+\sin^2{\theta} d\phi^2)\nonumber
\\&+4 P^2 \dfrac{\Delta}{\Sigma} (d\psi+\cos{\theta} d\phi)^2\,,
\end{eqnarray}
where
$$
    \Delta=\rho+3 M - \sqrt{M^2+2 P^2},\qquad \Sigma = \rho + M +
    \sqrt{M^2+2 P^2}.
$$
Here $\rho \geq 0$, the parameter $M$ is positive and the parameter
$P$ (usually called the magnetic charge) has the range $0<P\leq 2M$.
For $P<2M$ the metric (\ref{tnbh}) has the regular horizon at
$\rho=0$, so after \cite{gw} we shall refer to this solution as the
Kaluza-Klein black hole. For $P=2M$, the horizon degenerates to a
point and one gets the KK monopole (\ref{tn}) with $m=4M$.

Translating (12) into the ansatz (\ref{bcs}) we obtain
\begin{equation}\label{tnbh2}
    B=\frac{1}{3} \ln\left(\frac{\Sigma}{2P}\right), \, A =
    \frac{\rho r^2 (3\Sigma+\Delta)^2}{36 \Delta^2 \Sigma^3}, \,
    e^{2\delta}=\frac{\Delta}{\rho} A\,,
\end{equation}
where
\begin{equation}\label{rh}
    r = 4^{2/3} P^{1/3} \Sigma^{1/6} \Delta^{1/2}\,.
\end{equation}
Note that the leading order asymptotic behavior for large $r$ does
not depend on $M$ and is determined only by $P$
\begin{equation}\label{asym}
    A \sim \frac{4^3 P}{9 r},\qquad e^{2B} \sim \frac{r}{4P}\,.
\end{equation}
Although the KK black holes are known in closed form, for our
purposes it is helpful to look at them from the viewpoint of the
shooting procedure. Assuming that there is a non-degenerate horizon
at $r=r_H$, we obtain from (4) the following behavior
\begin{eqnarray}\label{local_h}
B(r)& \sim &\beta +
\frac{2(e^{-2\beta}-e^{-8\beta})}{4e^{-2\beta}-e^{-8\beta}}
\left(\frac{r}{r_H}-1\right), \nonumber \\ A(r) & \sim &
\frac{2}{3}(4e^{-2\beta}-e^{-8\beta})
\left(\frac{r}{r_H}-1\right)\,,
\end{eqnarray}
hence for each $r_H>0$ there is a one-parameter family of local
solutions with a regular horizon parametrized by $\beta=B(r_H)>0$.
Shooting these local solutions towards infinity we get
 the KK black holes (\ref{tnbh2}) where
\begin{eqnarray}\label{beta}
    \beta &= &\frac{1}{3} \ln{\frac{M+\sqrt{M^2+2P^2}}{2P}}, \\
    r_H^6 &= & 4^4 P^2
    (M+\sqrt{M^2+2P^2})(3M-\sqrt{M^2+2P^2})^3\,.\nonumber
\end{eqnarray}
\subsection{Large perturbations}
 Now, we return to the discussion of numerical results. We have
found that for sufficiently large perturbations the inner part of
the KK monopole collapses and a horizon forms at some
$r=r_H>0$.~Outside~the~horizon the solution settles down to the KK
black hole (see Fig.~3). As we pointed out above, the perturbations
considered by us have finite energy and therefore they do not change
the leading order asymptotic behavior (\ref{asym}). In this sense
the magnetic charge $P$ can be viewed as the constant of motion. For
the KK monopole $P=2M=m/2$, so all our configurations
 have the same magnetic charge $P=1/2$ (recall that we use units in which $m=1$)
  and the endstates of evolution differ only by
$M$.
\addtolength{\textheight}{.5in}  \addtolength{\footskip}{-.5in}
\begin{figure}[h]
\centering
\includegraphics[height=0.4\textwidth,width=0.48\textwidth]{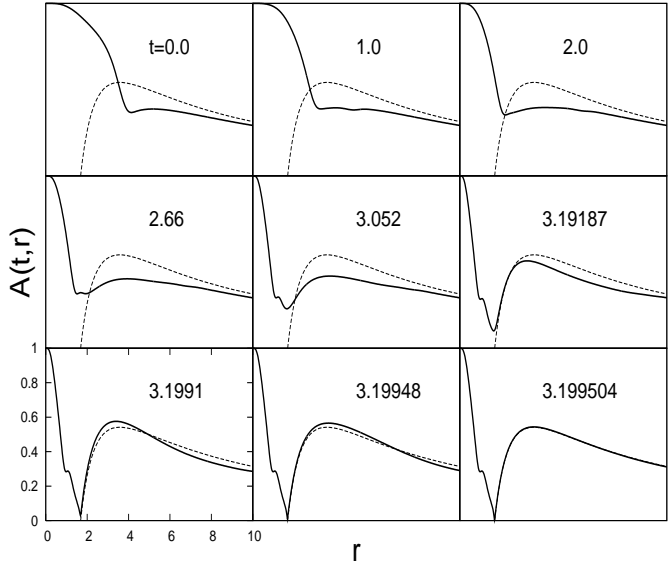}
\caption{\small{Instability of the Kaluza-Klein monopole for large
perturbations. For initial data  (\ref{idata}) with a large
amplitude ($p=0.3, R=3, s=1$) we plot a series of snapshots of the
function $A(t,r)$ (where $t$ is  central proper time). During the
evolution $A(t,r)$ drops to zero at $r=r_H \approx 1.68$ which
signals the formation of an apparent horizon there. Outside the
horizon the solution relaxes to the Kaluza-Klein black hole
(\ref{tnbh2}) (shown by the dashed line) with $M=0.502$. }
}\label{fig1}
\end{figure}

 As we approach the threshold of collapse from above, the parameter $M$
decreases towards the limiting value $M=1/4$ corresponding to the KK
monopole. At the same time, as follows from (\ref{beta}), both
$\beta$ and $r_H$ tend to zero. In this limit, close to the horizon
the KK black hole is very well approximated by the Schwarzschild
black hole.
 Near the threshold we observe the type II discretely
self-similar critical behavior governed by the same critical
solution as in the asymptotically flat situation \cite{bcs} (see
Fig.~4). This is not surprising in view of the fact that critical
collapse is a local phenomenon, and, as such, does not depend  on
the far field behavior.
\section{Conclusions}
 We have shown that although it is classically  stable against
small perturbations, the KK  monopole is classically unstable
against perturbations which are  large enough to allow gravitational
collapse to form a black hole. Classically, the black hole is
probably stable. Quantum mechanically one expects it to emit Hawking
radiation
 which can carry  off energy
but not magnetic charge. In the long run therefore one might expect
the system  to settle down to the original KK monopole. If this is
so, then the common intuition about BPS states would be correct,
provided one bears in mind  that the stability can only hold by
virtue of quantum mechanical effects.

\begin{figure}[tbh] \centering
\includegraphics[width=0.48\textwidth]{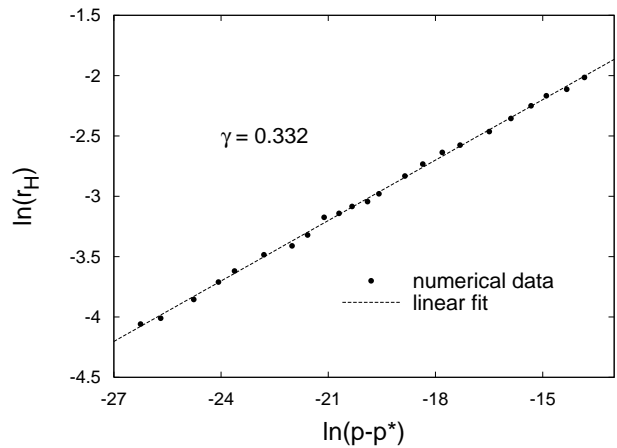}
\caption{\small{Critical behavior. For supercritical solutions we
plot the horizon radius vs. the distance to the threshold on the
log-log scale. The fit of the power law $\ln(r_H)=(\gamma/2)
\ln(p-p^*) + const$ yields $\gamma\approx 0.33$ which agrees (up to
numerical errors) with the critical exponent obtained in \cite{bcs}.
The echoing period $\Delta\approx 0.47$ (read off from the period of
 wiggles around the linear fit above and computed independently from the
spatial shift between the echoes) is also the same.}}\label{fig1}
\end{figure}

Finally, we remark that, besides the KK monopole, the BCS ansatz
incorporates other non-asymptotically flat solutions. In particular,
an exact regular time dependent solution is known~\cite{glp}
\begin{eqnarray}\label{timedep}
\!&ds^2\!=\!-dt^2\!\!+\!\!
\left(\dfrac{t}{m}+\dfrac{m}{\rho}\right)d\rho^2\! +\!\!
\left(\dfrac{t}{m}+\dfrac{m}{\rho}\right)\rho^2
(d\theta^2+\sin^2{\theta} d\phi^2) \nonumber \\
&+ m^2 \left(\dfrac{t}{m}+\dfrac{m}{\rho}\right)^{-1}
(d\psi+\cos{\theta} d\phi)^2\,.
\end{eqnarray}
This solution can be viewed as the time evolving KK monopole whose
$S^1$ fibers pinch off to a point for $t \rightarrow \infty$. It
would be interesting to determine the role of this solution in
dynamics.
\vskip 0.1cm {\textbf{Acknowledgments:}} PB acknowledges hospitality
of the Isaac Newton Institute in Cambridge during work on this
paper. The research of PB and TC was supported in part by the Polish
Ministry of Science grant no. 1~PO3B~012~29. 

\end{document}